\documentclass[12pt,draftcls,onecolumn]{IEEEtran}

\usepackage{epsfig,amsmath,amssymb,epsf,amsthm,scalefnt,multirow,subfig}
\usepackage{xcolor}
\usepackage{float}
\usepackage{cite}
\usepackage{psfrag}
\usepackage{booktabs}

\newtheorem*{lemma*}{Lemma}

\def\b0{{\pmb{0}}} 

\def\ba{{\mathbf{a}}}

   \def\bH{{\mathbf{H}}}

\begin{document}
%\bstctlcite{IEEEexample:BSTcontrol}

% paper title
\title{Millimeter Wave communication with out-of-band information}

 %author names and IEEE memberships
\author{Nuria Gonz\'{a}lez-Prelcic, Anum Ali, Vutha Va  and Robert W. Heath Jr.\\
\thanks{Nuria Gonzalez-Prelcic is with Universidade de Vigo, Vigo, Spain (email: nuria@gts.uvigo.es).}
\thanks{Anum Ali, Vutha Va and Robert Heath are with the Wireless Networking and Communications Group, The University of Texas at Austin, Austin, TX 78712, USA (email: \{anumali,vutha.va,rheath\}@utexas.edu).}
\thanks{This research was partially supported by the U.S. Department of Transportation through the Data-Supported Transportation
Operations and Planning (D-STOP) Tier 1 University Transportation
Center, by the Texas Department of Transportation under Project 0-6877 entitled Communications and Radar-Supported Transportation Operations and Planning (CAR-STOP), by the National Science Foundation under Grant NSF-CCF-1319556 and by the Spanish Government and the European Regional Development Fund (ERDF) under project MYRADA (TEC2016-75103-C2-2-R).}}

\maketitle

%\squeezeup\squeezeup\squeezeup\squeezeup
\begin{abstract}
Configuring the antenna arrays is the main source of overhead in millimeter wave (mmWave) communication systems. In high mobility scenarios, the problem is exacerbated, as achieving the highest rates requires frequent link reconfiguration. One solution is to exploit spatial congruence between signals at different frequency bands and extract mmWave channel parameters from side information obtained in another band. In this paper we propose the concept of out-of-band information aided mmWave communication. We analyze different strategies to leverage information derived from sensors or from other communication systems operating at sub-6 GHz bands to help configure the mmWave communication link. The overhead reductions that can be obtained when exploiting out-of-band information are characterized in a preliminary study.  Finally, the challenges associated with using out-of-band signals as a source of side information at mmWave  are analyzed in detail. 
\end{abstract}

\section{Introduction}\label{sec1}
%
%Paragraph 1: MmWave communication using large arrays. Small antennas posible. Potential applications
The use of  mmWave spectrum provides
access to high bandwidth communication channels, leading to gigabit-per-second data rates.
MIMO systems operating at these frequencies need large arrays that provide enough antenna aperture and array gain. 
Antenna arrays with a large number of elements are small at mmWave, which enables its potential use in applications where the size and weight of the radio-frequency stage is a limiting factor, such as wearable networks, mobile devices or virtual reality  devices. For example, \cite{HonBaeLee:Study-and-prototyping-of-practically:14} describes a prototype  mmWave cellular phone developed by Samsung equipped with two sets of 16-element antenna arrays on the top and bottom parts of the smart phone board.  
%In \cite{ChaZhaMuh:60-GHz-Textile-Antenna:13}, a commercial textile is used as substrate for a microstrip patch antenna array that can be used in body-centric short range communications at 60 GHz. 
%(not sure it is a good example, only 4 patches, see review paper in the folder for a better reference)

%Paragraph 2: Benefits of large arrays: MU-MIMO possible at mmWave; less interference, rate multiplication; better spectrum shating;
The potential benefits of large mmWave arrays are not limited to the array gain they provide or to their small weight and size.
Large arrays enable highly directional transmission and reception, which reduces the amount of interference in the mmWave communication system and contributes an additional  gain in data rates.
%as shown in \cite{BaiHea:Coverage-and-Rate-Analysis:15}. 
Better spectrum sharing between cellular operators is also possible at mmWave. When using narrow beams, the per-user rate increases  when sharing spectrum, even if there is no coordination between the operators \cite{GupAndHea:On-the-Feasibility-of-Sharing-Spectrum:16}.

%Paragraph 3: Challenges of using large arrays. Main one is configuring the arrays. Review beam training and compressed channel estimation strategies. Still too much overhead.
The benefits of using mmWave large arrays do not come for free.  The main challenge when using mmWave is minimizing the link establishment and operation overhead. Several strategies  for configuring the links  when using different kinds of MIMO architectures have been proposed in the previous literature \cite{HeaGonRan:An-Overview-of-Signal-Processing:16}. Analog beamformers in conventional single-stream mmWave are designed by following a closed-loop beam-training strategy based on searching  a codebook which includes beam patterns at different
resolutions. Though this approach leads to the selection of the best transmit-receive beam pair, it usually involves high overhead.
Hybrid analog-digital beamformers are normally configured from the estimate of the MIMO channel matrix. Compressive channel estimation algorithms which exploit sparsity in the mmWave channel have been proposed as an efficient tool to acquire channel information. The overhead of this approach is lower than the one corresponding to beam training protocols, but still too high in high mobility scenarios, which require frequent beam reconfiguration.

\begin{figure}
\centering
{
\includegraphics[width=0.7\columnwidth]{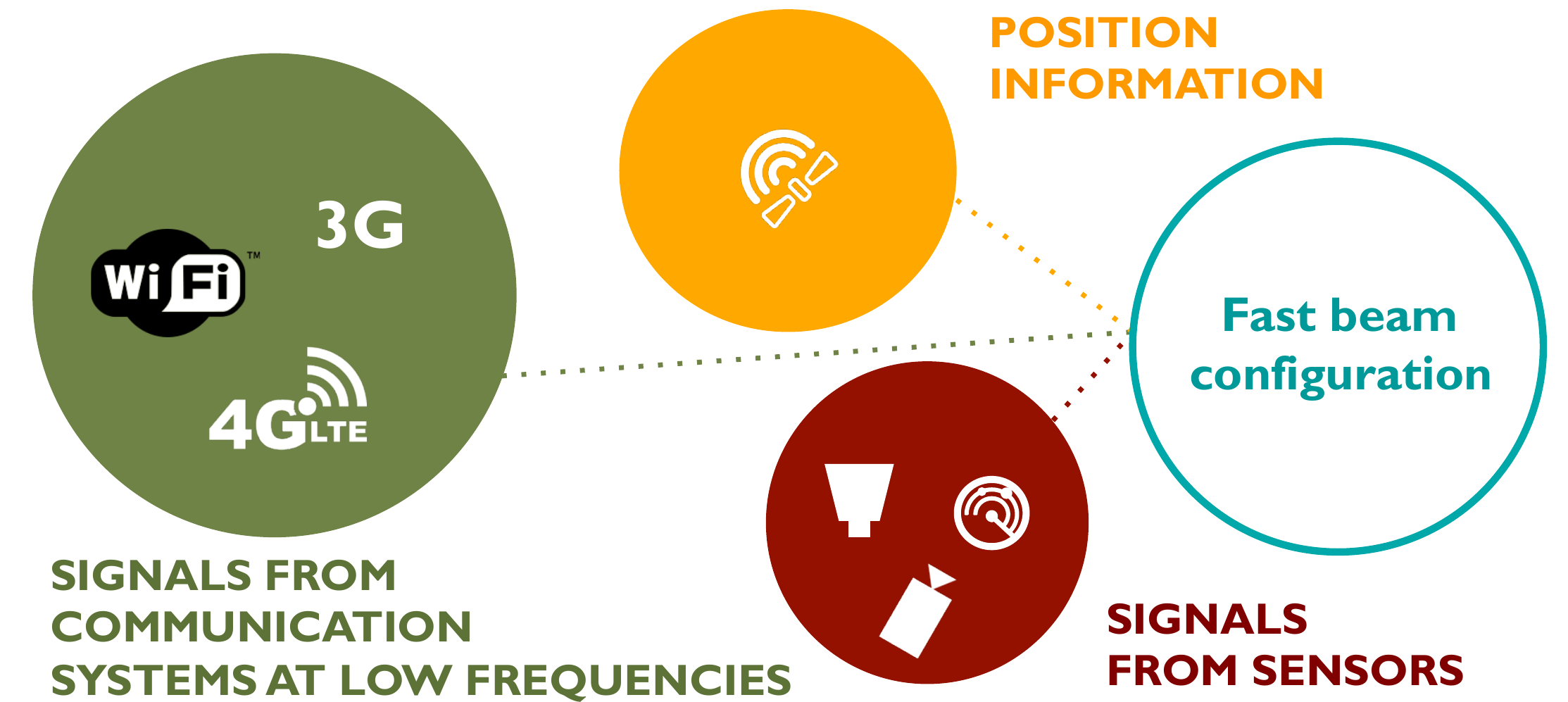}
\caption{Sources of out-of-band information to aid mmWave link establishment.}
\label{OOBinfo_sources}
}
\end{figure}

%Paragraph 4: Fast beam configuration using out of band information coming from sensors or other communication systems. Figure from slide 8 in the keynote.
In this paper, we propose to use out-of-band information coming from sensors or other communication systems operating at sub-6GHz frequencies to aid 
mmWave communication link establishment, significantly reducing the associated overhead. The key idea, illustrated in Fig. \ref{OOBinfo_sources}, is using information extracted from one frequency band as a prior to acquire channel information in another band. The main motivation for using lower frequency signals is that  mmWave cellular  systems will likely be deployed in conjunction with sub-6GHz systems. Consequently, the corresponding communication signals can be used to obtain coarse estimates of channel parameters without taxing the mmWave communication resources. Further, sensors are everywhere nowadays, e.g. automotive sensors or GPS in a mobile device, or can be easily installed in a mmWave base station if needed. These sensors can provide information about the environment that can be processed as well to infer some channel characteristics, e.g. direction of arrival from a radar signal. Our vision is that out-of-band information will be the key that unlocks the potential of mmWave communications in high mobility scenarios.

\section{Going out-of-band}

%Paragraph 1: Is this a crazy idea? Example of channel parameters reciprocity in FDD between close bands.
The concept of exploiting information from one band to aid communication in another band is an old idea. Most cellular communication systems use slightly different carriers for the transmitter
and receiver, meaning that the forward channel (from base station to mobile station) differs from the backward
channel (from mobile station to base station). This motivates the use of limited feedback to inform
the base station about the forward channel. An alternative is to exploit reciprocity between channel parameters, and to perform beamforming
based on the spatial covariance matrix  and statistical reciprocity. While the 
channel itself is not reciprocal, prior work has argued that the spatial covariance matrix is approximately
reciprocal \cite{AstForFet:Downlink-beamforming-avoiding:98}, since forward and reverse communication occurs through the same propagation paths. 
Though this is an interesting idea, it exploits similarities between very close bands, while we are interested in
exploiting the similarities between sub-6 GHz and mmWave channels.
%Not sure if it is interesting to include more details and equations here about the relationship RL and RH

%Paragraph 2: Describe propagation differences between sub-6 GHz and mmWave. How similar are the spatial characteristics of different bands?
The propagation characteristics of mmWave signals are different, which leads to mmWave wireless channels with a different structure.
Signals at mmWave frequencies are more sensitive to blockage, because millimeter waves do not penetrate solid materials very well.
Due to the small wavelength, diffraction is not significant at mmWave, and the reflections often less specular.
Though more scatterers appear, only a few paths with enough energy exist. This together with the large sizes of the mmWave arrays
lead to a sparsity assumption on the mmWave MIMO channel matrix, which can be exploited for beam training, 
channel estimation or precoder/combiner design. It is clear that  communication channels at sub-6 GHz and mmWave 
do not share the same parameters, and in some cases can not even be modeled using the same approach, 
but maybe there exists some similarity between spatial characteristics (directions of arrival or departure, angle spread, etc.)
that can be exploited.

%Paragraph 3: Describe results on spatial characteristics across bands separated on the order of few MHz and between channel parameters separated on the order of several GHz.
Early work on FDD channel reciprocity established the spatial similarity
between channels separated by a few megahertz. Measurements reported in \cite{HugKalLau:Spatial-reciprocity-of-uplink:02} showed that the DoAs of dominant paths for channels at  1935MHz and 2125 MHz are very similar with high probability. Surpringsingly, recent measurements confirm that some channel parameters are also similar when the center frequencies are separated by tens of gigahertzs. For example, multi-band channel sounding performed in \cite{Petal.:Measurement-campaigns-and-initial:16} showed almost identical spatial characteristics at 5.8, 14.8 and 58.7 GHz, and minor differences in the cumulative distribution functions (CDF)  of DoA/DoD or delay spread were observed in \cite{KyCarKar:Frequency-dependency-of-channel:16} when measuring the channel at six frequencies between 2 and 60 GHz.

%Paragraph 5: Describe measurement results when comparing sub-6 GHz and mmWave bands.
Initial work that proposes the use of WiFi signals at  sub-6 GHz to aid the configuration of 60 GHz WiFi links   \cite{NitFloKni:Steering-with-eyes:15} reports some measurements that confirm the spatial congruence between sub-6 GHz and mmWave channels. Though more multi-band measurement campaigns are needed to establish a more precise relationship between spatial characteristics of sub-6 GHz and mmWave channels, there is enough evidence to propose the use of side spatial information obtained from sub-6 GHz signals as coarse estimation of the spatial parameters at mmWave. In the rem
ainder of this paper, we assume there is enough spatial congruence between these  vastly separated channels. First, we review the challenges associated to the translation of the spatial information. Second, we describe several approaches to exploit different types of side information. Finally, we evaluate the overhead reduction that can be obtained when using out-of-band information to aid configuring the mmWave arrays.

\section{Challenges of translating channel information between vastly different bands}\label{sec:cv_translation}

%Paragraph 1: Channel parameters of interest. DOA, angle spread, covariance. Beamforming based on these paramters. How to translate these parameters from one band to another one. 
The design of optimal precoders/combiners to maximize the achievable rate needs perfect channel knowledge. There are other metrics though, such as MSE or average SINR, which lead to solutions for the precoders and combiners that only depend on estimations of the
spatial autocorrelation of the received signal and the cross-correlation of the transmitted and received signal. These designs are interesting because 
the training overhead can be greatly reduced if the channel does not have to be perfectly estimated.
Designing this kind of precoders from out-of-band information requires spatial correlation translation techniques, which can provide an estimate of the spatial correlation at mmWave from an estimate obtained from a sub-6 GHz signal.

%Paragraph 2: Review previous work on correlation translations for SISO. Limitations of this work. Our results on SIMO. Limitations.
In the context of exploiting channel reciprocity for FDD systems, prior work proposes different strategies 
to translate spatial correlation. Some work \cite{AstForFet:Downlink-beamforming-avoiding:98} proposes to use least squares to find a transformation operator that converts the steering vectors at one frequency to steering vectors at another frequency. Spline interpolation has also been used in previous work on spatial correlation translation between the uplink and the downlink  in FDD systems \cite{JorGonAsc:Conversion-of-the-Spatio-Temporal-Correlation:09}.

\begin{figure}
\centering
{
\includegraphics[width=0.7\columnwidth]{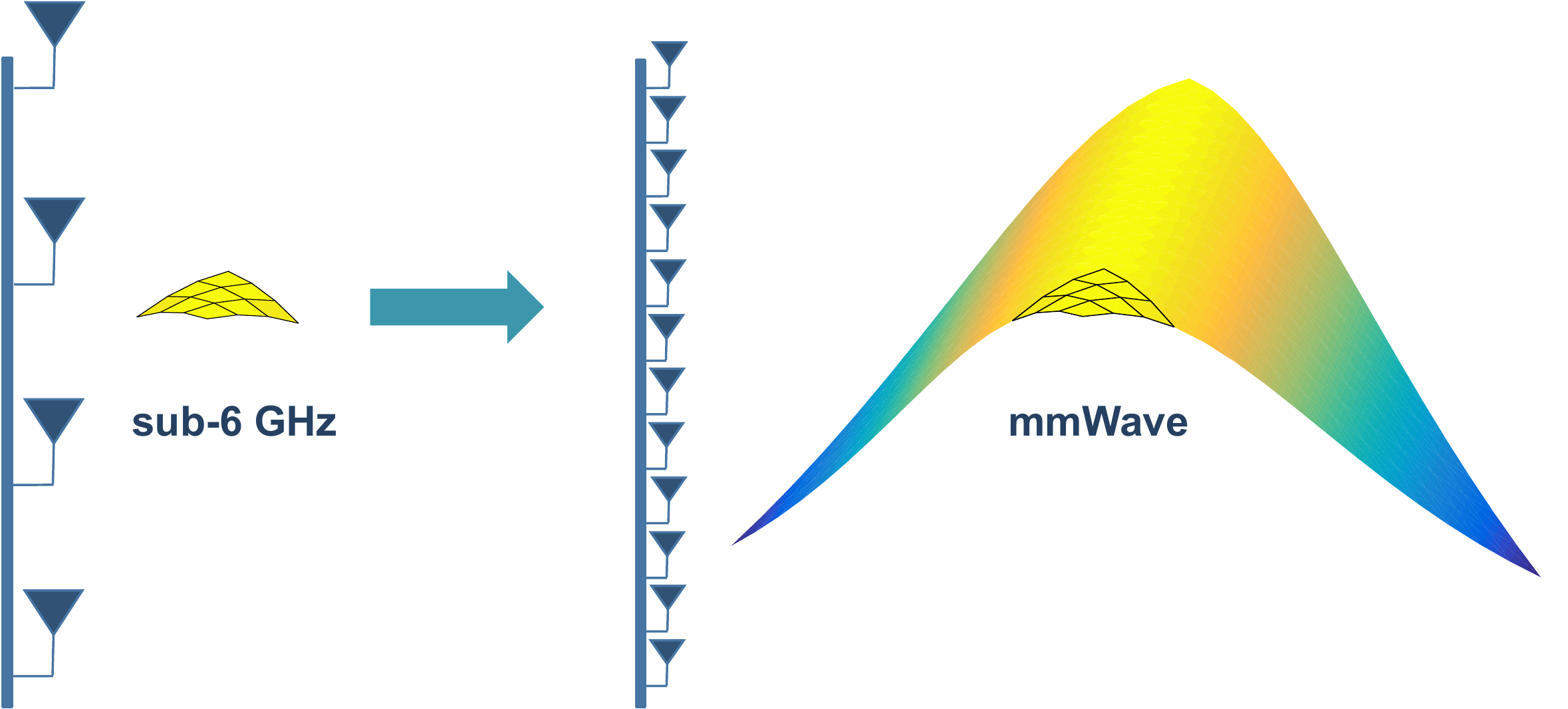}
\caption{Representation of the spatial correlation  matrix for a sub-6GHz and a mmWave MIMO system operating with the same antenna aperture. The spatial correlation at mmWave has a much larger size, since a much larger antenna array is needed at high frequencies.}
\label{cov_translation}
}
\end{figure}

%Paragraph 3: Comment on challenges related to correlation translation in MIMO systems. Size of correlation matrices. Slightly different received power, angle spread and DOAs. Any prior work on translating other parameters?
Translating the correlation information from sub-6 GHz to mmWave poses additional challenges. The correlation matrices at low and high frequencies differ in size because the number of antenna elements used in the antenna arrays at sub-6 GHz and mmWave are different in the same aperture. In addition, there is a mismatch in the angles-of-arrival and the angle spread. The transformation from a smaller to a larger dimension is particularly challenging, as shown in Fig. \ref{cov_translation}, since a large correlation matrix has to be estimated from low-frequency data obtained at a few antennas.
Recent work  \cite{AliGonHea.-:Estimating-millimeter-wave:16} proposes a non-parametric and a parametric spatial correlation translation technique  for SIMO systems using comparable apertures at mmWave and sub-6 GHz frequencies. The non-parametric approach exploits the structure in the spatial correlation matrix, and uses interpolation/extrapolation to obtain the high frequency correlation from the low frequency one. In the parametric approach, theoretical expressions of the high frequency correlation are used. The parametric approach performs better but requires knowledge of the antenna array geometry used at the receiver side and the distribution of DoAs.
Though this initial work gives some insights on the design of correlation translation strategies between sub-6 GHz and mmWave, additional work is needed
to fully exploit out-of-band information at mmWave: considering other array geometries when using parametric translation approaches, accounting for channel parameters mismatch, or extending the results to the MIMO case.

\section{Exploiting out-of-band information}
%Paragraph 1:  Summarizing the section and explaining the different types of OOB info to be considered.
In this section, we propose to leverage different types of signals as side information to extract coarse estimates of different channel parameters  at mmWave.
We consider signals from sub-6 GHz communication systems and also raw or processed sensor data, and analyze different approaches to extract prior information that aids mmWave beam training. The challenges associated with each approach are also analyzed.

\subsection{Exploiting position information}\label{sec:position}

%Paragraph 1: Different sources of position information. How to exploit position info to restrict the set of candidate beams.
There are several opportunities to exploit position information to help
reduce the beam alignment overhead at mmWave. Position information can be obtained from GPS at the user side, or cameras or radar at the BS side.
In a vehicular communication context, position information can also be extracted from vehicular communication systems operating at lower frequencies such as DSRC, in which cars broadcast their positions periodically. Under the line-of-sight (LOS) dominant scenario,  position information can be used to compute the pointing direction
and essentially eliminate the beam alignment overhead. It is important to consider the effect of position estimation error in the  choice  of beamwidth. It was shown in \cite{VaZhaHea:Beam-Switching-for-Millimeter:15} that there exists a trade-off in the beam width selection: very narrow beams are sensitive to position error, while wide beams provide a low antenna gain. Given a model for the position error, optimal beam designs can be found. We conclude that position information is a valuable option for fast beam configuration.

% Fingerprinting
There are other alternatives to obtain position information, for example multipath  fingerprinting. This is a technique used for indoor localization. It is based on measuring the multipath signature to compare it  to a fingerprint database and
obtain the most likely position. This procedure can also be applied in reverse: the infrastructure can collect  a database of multipath fingerprints, i.e. DoA/DoDs, of paths indexed by location, and then take an estimate of position to generate a multipath angle-delay profile. This approach has been recently proposed in \cite{VaChoShi:Inverse-Fingerprinting-for-Millimeter:17} for mmWave vehicle-to-infrastructure (V2I) beam alignment. In Section \ref{sec:evaluation} we analyze the overhead reduction provided by inverse fingerprinting when compared to the beam training protocol in IEEE 802.11ad.

%Paragraph : Summary of challenges
Recent work has shown that position information can provide fast mmWave beam alignment in V2I scenarios, but additional work is needed to
extend the approach to the cellular case with mobile handsets. In this scenario, position information could be combined with device orientation to 
obtain appropriate beam alignment strategies. Position information could also be combined with other out-of-band information like velocity to improve accuracy in the beam alignment process. Robustness to dynamic blockages in the environment also has to be incorporated in the design of position-based beam alignment strategies.

\subsection{Exploiting communication signals at sub-6 GHz frequencies}

%Paragraph 1: Vision of future cellular networks with multiband connectivity. Other communication signals possibly available coming from WiFi networks for example. Motivation for using OOB in this case. 
Multi-band connectivity, where both mmWave and conventional sub-6 GHz frequencies are supported, is a likely feature of next generation cellular networks. 
Other communication signals such as lower frequency WiFi signals  may be possibly available around a mmWave communication system and have colocated infrastructure.
These systems operating in parallel can also be a good source of free  side information to aid in the establishment of the mmWave link.
  
 \begin{figure}
\centering
{
\begin{tabular}{cc}
\includegraphics[width=0.45\columnwidth]{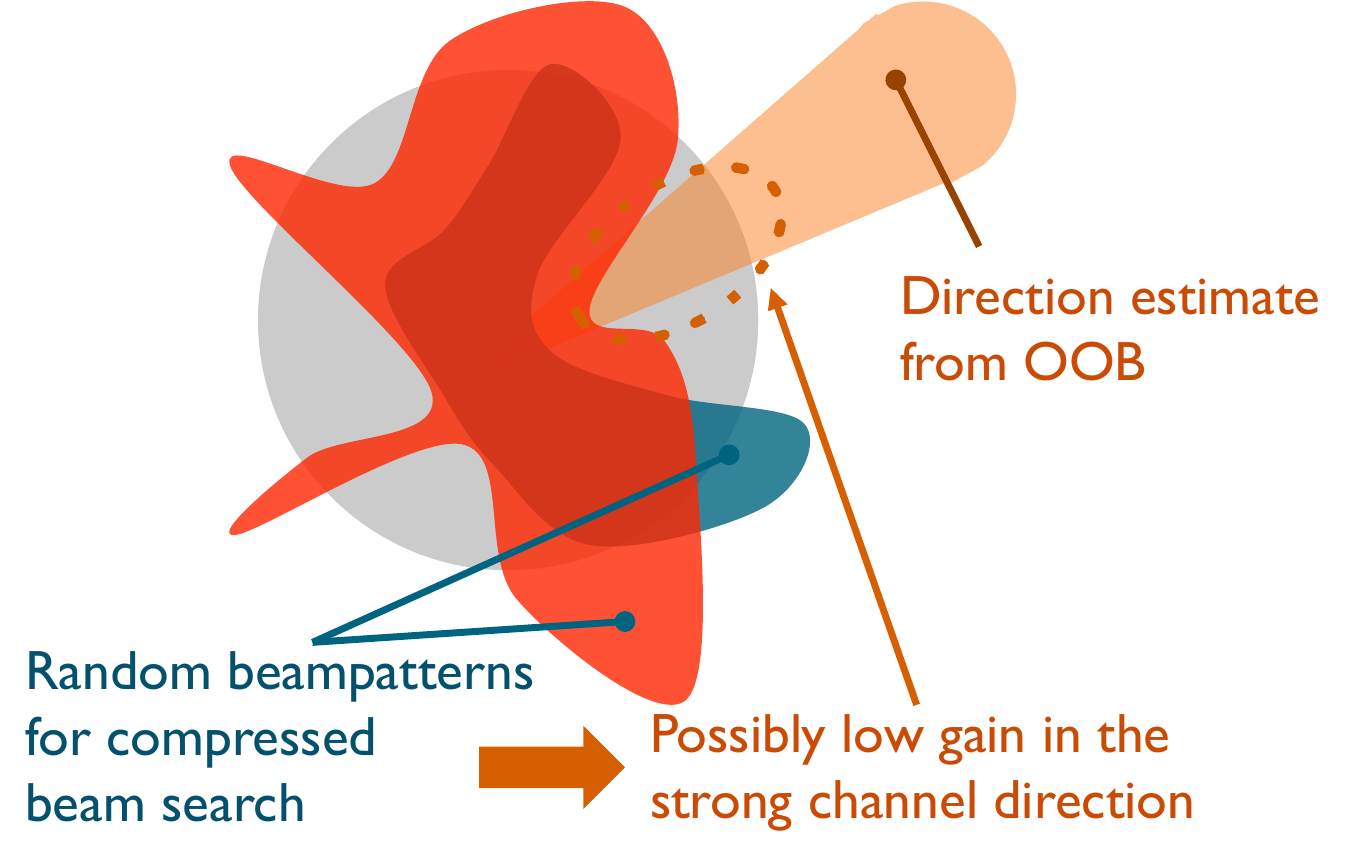} & \includegraphics[width=0.37\columnwidth]{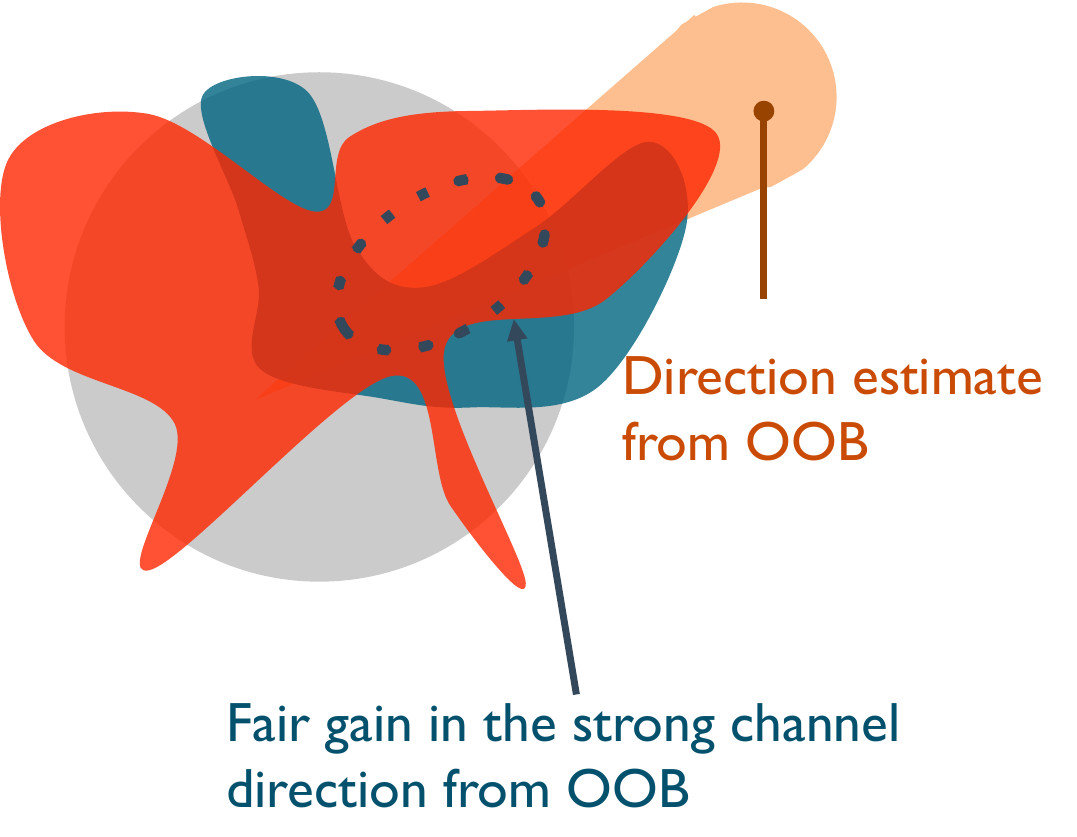} \\
(a) & (b) \\
\end{tabular}
\caption{(a) Dictionary of random beam patterns used in compressive beam search approaches. (b) Structured random dictionary with  random beams shaped using out-of-band information.}
\label{fig:structured_random_dictionaries}
}
\end{figure}
 
%Paragraph 2: Precoding based on covariance information. Link to section 3. Beam search with OOB information.
One approach to use this type of side information could be designing the precoders and combiners based on the spatial correlation, estimated
from a lower frequency communication signal, as discussed in Section \ref{sec:cv_translation}.
Other  directions are also possible. A result on beam training exploiting out-of-band information \cite{AliGonHea:Millimeter-Wave-Beam-Selection:17}
formulates the beam selection as a weighted sparse recovery problem. The weights are obtained from angles of arrival/departure estimated from 
a sub-6 GHz signal, by penalizing the directions of the beams that are likely to be zero. The mismatch between mmWave and sub-6 GHz paths is also
incorporated when designing these weights. 
A different idea illustrated in Fig. \ref{fig:structured_random_dictionaries} consists of using out-of-band information to design structured random dictionaries of beam patterns, instead of the purely random approach. Out-of-band information shapes the beams in the dictionary for better compressive estimation of the channel, guaranteeing a fair gain in the strong channel directions obtained from out-of-band information.

%Paragraph 4: Hierarchical beam search  with OOB
Out-of-band information can also be used in hierarchical beam training protocols. Out-of-band aided hierarchical beam search may replace the coarse beam training stage by a direction estimate obtained from lower frequencies. This way, the number of required stages in the beam training process can be reduced. Further,  the noise level will likely be lower at sub-6GHz, so the wrong decisions on the strongest sector in the coarse estimation stage which lead to a failure in the beam training process are avoided. 

%Paragraph 5. Summary of challenges
Developing integrated protocols that allow joint training at low and high frequencies is an interesting challenge. The idea of exploiting out-of-band information not only for beam search but for channel estimation  has not been investigated yet. 

%\begin{figure}
%\centering
%{
%\begin{tabular}{ccc}
%\includegraphics[width=0.35\columnwidth]{figs/sensing_BS.pdf} & \includegraphics[width=0.2\columnwidth]{figs/sensors_mobile.pdf} &  \includegraphics[width=0.3\columnwidth]{figs/sensors_car.pdf} \\
%(a) & (b) & (c) \\
%\end{tabular}
%\caption{(a) Sensing mmWave base station equipped with radar, cameras or lidar. (b)  Mobile device including sensors such as GPS or accelerometers. (c) Vehicle equipped with many types of automotive sensors such as lidar, radar, cameras or GPS.}
%\label{sensor_types}
%}
%\end{figure}

\subsection{Exploiting signals form sensors}

%Paragraph 1: Types of sensors that can be exploited. Sensors at the BS. Sensors at the MS. Automotive sensors, etc.
Sensors are being  rapidly integrated in all the different electronic systems we use, in our environment and even in the small personal devices we wear everyday.
For example, 
%as shown in Fig. \ref{sensor_types}, 
automotive sensors like radar or cameras are becoming an essential ingredient of new vehicles, with more and more automation capabilities; our smart phones are already equipped with GPS, gyroscopes, or  accelerometers, and future base stations can be easily equipped with different types of sensing technologies to acquire information about the environment. Sensors are already everywhere, and can provide information for free that can be useful for a mmWave communication system.

\begin{figure}
\centerline{\includegraphics[width=0.7\columnwidth]{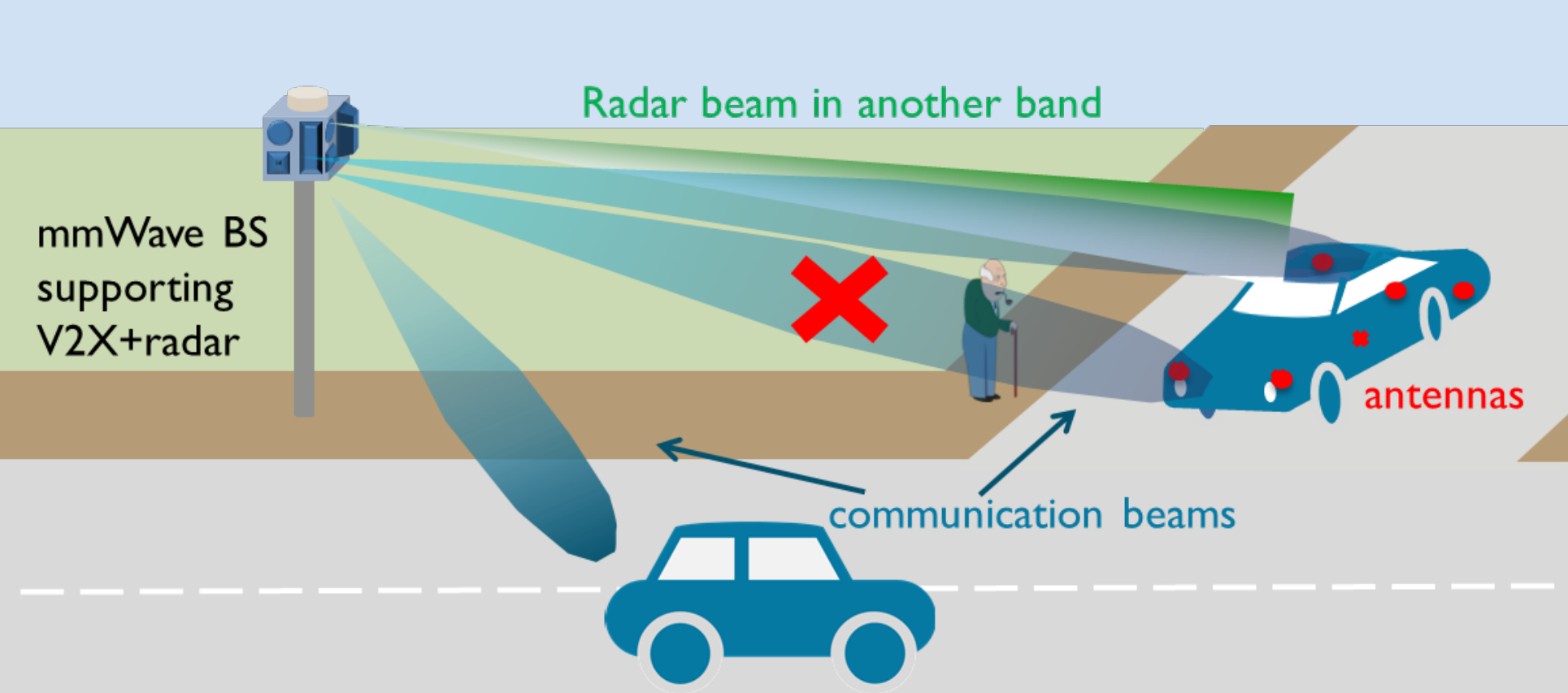}}
\caption{Illustration of a radar-aided mmWave vehicular-to-infrastructure communication system.}
\label{fig:radar_aided_commun}
\end{figure}

%Paragraph 2: Radar aided mmWave communication. Explain the concept and preliminary results in ITA paper.
The concept of sensor aided vehicular communications at mmWave is new. A radar mounted on a base
station can capture a large view of the surrounding area and help the infrastructure with
multiuser communication.  
Radar can be used to  reduce overhead in the first coarse configuration  of the beam directions,
what  can be followed by a second phase of digital hybrid precoder optimization if a hybrid architecture is being used,
or a beam refinement stage in analog-only MIMO architectures. An initial approach \cite{GonMenHea:Radar-Aided-Beam:16}
uses ray tracing  simulation to show that most of the dominant DoAs for the mmWave communication signal also appear at the 
radar echo in a different mmWave band. Then a multiuser beamformer is designed from the spatial correlation of the radar echo,
considered a coarse estimation of the spatial correlation of the true communication signal. This idea, applied in a vehicle-to-infrastructure (V2I) communication system is illustrated in Fig.~\ref{fig:radar_aided_commun}.

%Paragraph 3: Machine learning to process sensors signals and predict channel parameters.
Sensing at the infrastructure can also be used for for blockage prediction. In mmWave links,
vehicles and pedestrians may block the primary (usually LOS) communication path. Static objects in the
environment, e.g., trees and buildings are another source of blockage. Assuming that the base station is equipped with a radar, 
 a combination of radar and machine learning  can detect potential obstacles and their associated mobility, to improve mmWave communication performance.  Information derived from the radar echo (dominant sources of scattering,
velocity of scatterers, etc) will be a source of features. Past communication performance will be exploited
in the machine learning algorithm to classify particular radar responses as blockages. Combined with the
map of the static environment, this information can be used to develop an algorithm for the prediction of the
different types of blockage that a mobile station may experience. The results of the blockage
prediction algorithm can be used to redefine the new beams that have to be used at the infrastructure side to
illuminate the users. 
 
%Paragraph 4: Challenges on exploiting sensor information.
Many research challenges remain unsolved when exploiting sensing information to aid mmWave communication. 
A clear relationship of the spatial congruence between radar and communications has to be established from measurements. 
This way, it will be possible to compute transformations of the radar correlation which will be better estimates of the communication signal correlation.
Simultaneous sensor and communication measurements are also needed to develop similarity models.
Other sensors such as lidars or cameras can be considered as well, and how to extract channel parameters from their corresponding
raw signals has not been studied in the literature. Machine learning algorithms that can combine sensor and communication information to gain a priori knowledge of the mmWave channel is also a key topic to be explored. 

\section{Evaluation of out-of-band information aided mmWave communication}\label{sec:evaluation}

The main advantage of using out-of-band information to aid configuring the mmWave link is overhead reduction, or seen from another perspective, an improvement in performance when using a fixed channel training length. In this section, we discuss some examples of using out-of-band information and their associated benefits.

The work in \cite{AliGonHea:Millimeter-Wave-Beam-Selection:17} presents some specific algorithms for beam search at mmWave exploiting out-of-band information extracted from a sub-6 GHz communication system. The first algorithm, weighted basis pursuit denoising (W-BPDN), solves a compressed  beam search problem weighting the support of the channel according to the prior information provided by out-of-band measurements. This way, the entries that are likely to be zero are penalized during the search process. A second approach, structured weighted basis pursuit denoising (SW-BPDN), performs the first algorithm over structured random dictionaries, which 
contain training beams with a  gain over a given threshold in the angular directions suggested by out-of-band information.
We include here some simulations to illustrate the gains of these approaches.

\begin{table}
\begin{center}
\begin{tabular}{ | l | c | c |}
    \hline
    & {\bf Sub-6 GHz } & {\bf MmWave} \\ \hline \hline
    Carrier frequency & 3.5 GHz & 28 GHz \\ \hline
    Bandwidth & 1 MHz & 320 MHz \\ \hline
    Number of antennas at the TX & 4 & 32 \\ \hline
    Number of antennas at the RX & 4 & 32 \\ \hline
    Array geometry & ULA, $d=\lambda/2$ & ULA, $d=\lambda/2$ \\ \hline
    Phase shifter resolution & N/A & 5 bits \\ \hline
    Transmit power & 37 dBm & 37 dBm \\ \hline
    Path loss coefficient & 3 & 3 \\ \hline \hline
   \end{tabular}
   \caption{Simulation parameters for the sub-6 GHz aided mmWave compressed beam search.}
   \label{tab:setup_one}
   \end{center}
   \end{table}

\begin{figure}
 \begin{tabular}{cc}
\includegraphics[width=0.455\columnwidth]{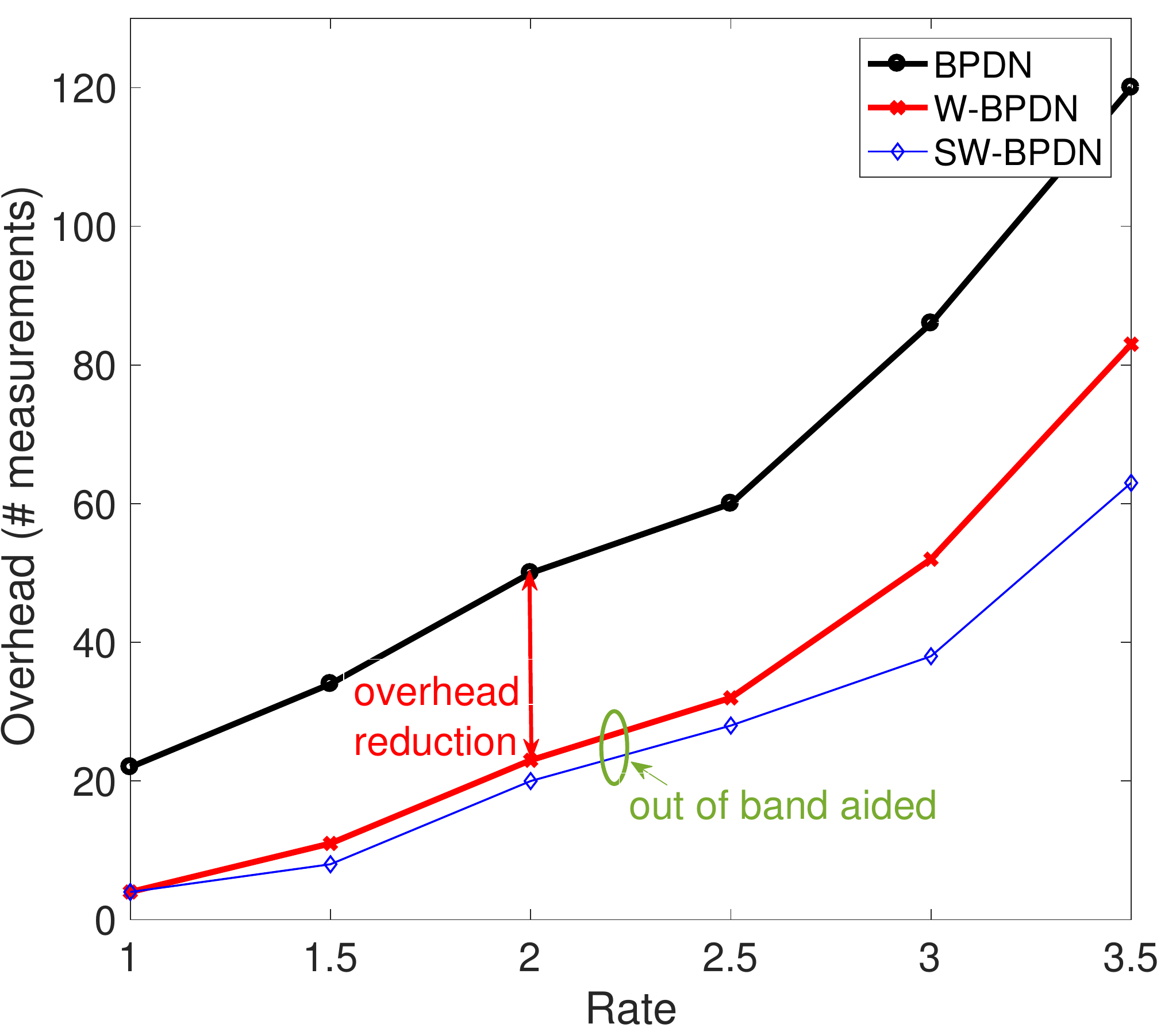} & \includegraphics[width=0.5\columnwidth]{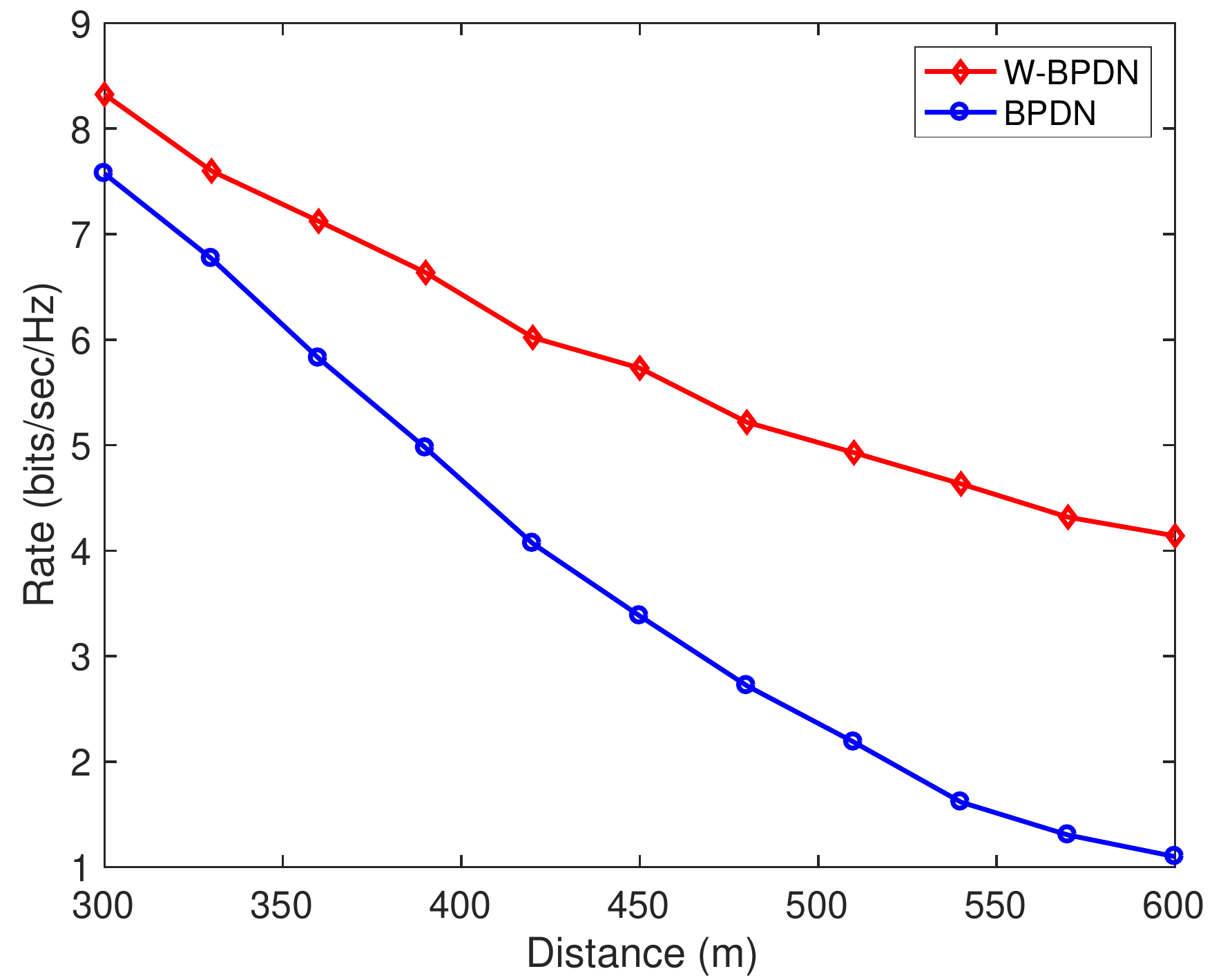}\\
(a) & (b) \\
\end{tabular}
\caption{(a) Overhead reduction provided by W-BPDN and SW-BPDN when compared to an equivalent beam search algorithm which does not exploit 
out-of-band information when the distance between transmitter and receiver is 40 m. (b) Effective rate for different beam search strategies exploiting out-of-band information as a function of the distance between the transmitter and the receiver.}
\label{fig:OOB_overhead_reduction}
\end{figure}

We consider first the simulation setup in Table~\ref{tab:setup_one} to evaluate the effectiveness of using a sub-6 GHz communication system to aid beam configuration in a mmWave link. We use a narrowband geometric channel model for the sub-6 GHz system, and a frequency selective wideband mmWave channel model with 63 taps.  For the transmission we consider OFDM, with 256 subcarriers and a cyclic prefix length of 64. A raised cosine filter with roll-off 1 is assumed to model the convolution of the pulse shaping filter and the matched filter. 
Fig.~\ref{fig:OOB_overhead_reduction}(a) shows the overhead reduction provided by W-BPDN and SW-BPDN with respect to an equivalent beam search process which does not exploit out-of-band information when the transmitter and receiver are separated by 40 m. Another way to see the benefits of out-of-band information is analyzing the effective rate provided by the out-of-band information aided approach and the standard basis pursuit denoting (BPDN) approach when the training length is fixed. Fig.~\ref{fig:OOB_overhead_reduction}(b) shows
the effective rate as a function of the distance between the transmitter and the receiver for the previous setup, when the number of measurements is fixed to 36.
The effective rate is defined as $R_\text{eff}=\eta\log_2(1+\vert \ba^*_{\text{RX}}(\theta_{\hat{n}})\bH \ba_{\text{TX}}(\phi_{\hat{n}})\vert^2 SNR)$,
where $\eta=\max(0,1-(N_{\text{TX}}\times N_{\text{RX}})/L_\bH)$ is the loss due to training, with $/L_\bH$ the channel coherence time, $\ba^*_{\text{RX}}$ and $\ba_{\text{TX}}$ are the steering vectors at the receiver and transmitter, and  $\hat{n}$ and $\hat{m}$ the estimated transmit and received codeword indices. The gain provided by out-of-band information increases with the distance, because the SNR decreases and the conventional beam search is not effective any more.

\begin{figure}
\begin{tabular}{cc}
\includegraphics[width=0.5\columnwidth]{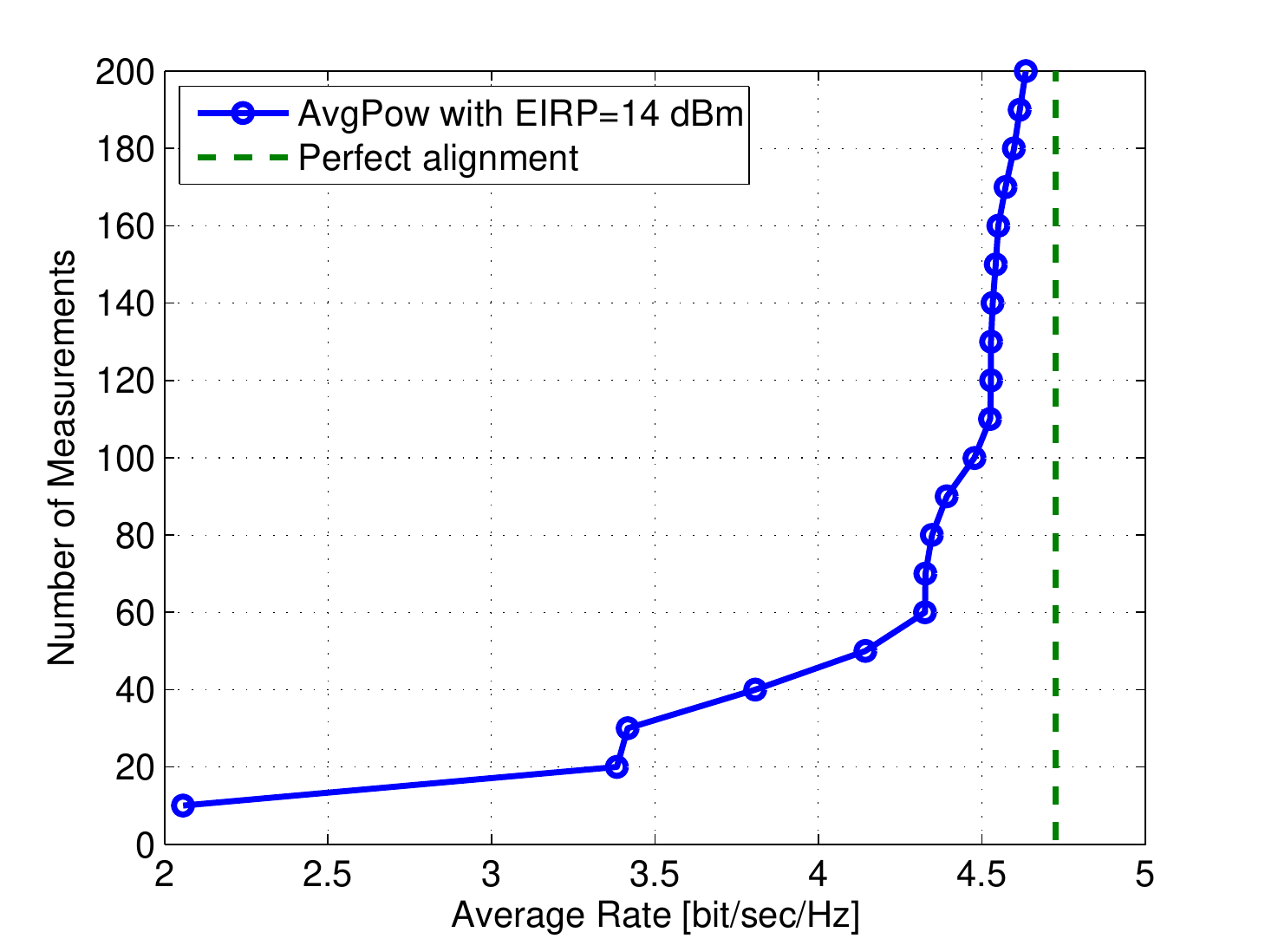} & \includegraphics[width=0.5\columnwidth]{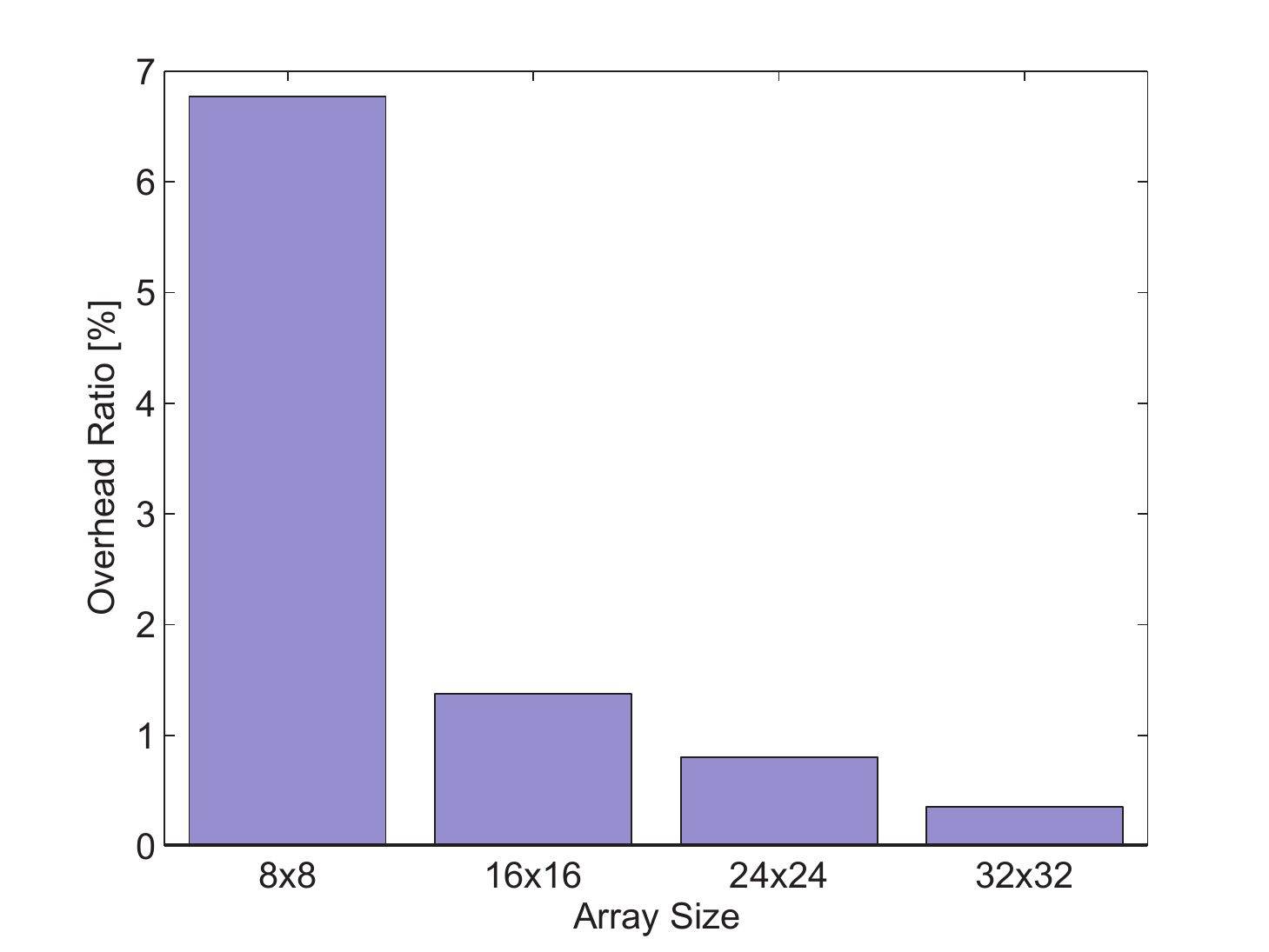} \\
(a) & (b) \\
\end{tabular}
\caption{(a) Overhead for the inverse fingerprinting approach as a function of the average rate. (b) Overhead of fingerprinting as a percentage of the overhead of IEEE 802.11ad versus the array size.}
\label{fig:overhead_fingerprinting}
\end{figure}

As an example of using position information to reduce overhead we can consider the inverse fingerprinting approach discussed in Section \ref{sec:position}.
To evaluate the overhead of this approach, we define the power loss  as the difference in received power in dB when operating with the optimal beam pair computed from exhaustive search and the pair selected by inverse  fingerprinting. Now the overhead is defined as the number of beam pairs to be trained such that the probability of power loss is less than 3 dB is $>99\%$ multiplied by the number of symbols used to train each beam. 
To evaluate the performance of this approach we consider a ray tracing simulation setup, assuming uniform planar arrays of sizes  $8\times8$, $16\times16$, $24\times24$ and $32\times32$ with isotropic antenna elements.
The curve in Fig.~\ref{fig:overhead_fingerprinting}(a) shows the evolution of the absolute overhead of the inverse fingerprinting approach necessary to achieve a given average rate when the SNR is -16.88 dB, each beam is trained using 10 OFDM symbols and the array sizes at the transmitter and receiver are $16\times16$ uniform planar arrays. It can be clearly observed that the proposed beam alignment scheme can achieve close to perfect alignment rate with small training overhead.    Fig.~\ref{fig:overhead_fingerprinting}(b) shows the overhead of fingerprinting as a percentage of the overhead of IEEE 802.11ad as a function of the array size. To compute the IEEE 802.11ad beam training overhead we assume a two-stage beam training with $N_{QO}$ narrow quasi-omni patterns and $N_\text{sec}$ sector patterns,  using $32$ spreading for sector search. Training length at the quasi-omni should be 32 times longer than at the sector level to achieve the same SNR level, and $N_{QO}=\frac{N_\text{sec}}{32}$.
This way, the overhead for the IEEE 802.11ad protocol is computed as $T_\text{11ad}=N_\text{QO}^2\times 32T_\text{tr}+2\frac{N_\text{sec}}{N_\text{QO}}T_\text{tr}$, where $T_\text{tr}$ is the training length at sector level. 
This expression can also be written as a function of the number of antennas at the transmitter and receiver,  $T_\text{11ad}=\left(\frac{N_a^2}{32}+64\right)T_\text{tr}$.
Note that IEEE 802.11ad overhead is quadratic in the number of antennas, while
 the overhead of inverse fingerprinting depends on the angle spread, and
the number of beams covering a fixed angular area, so that it increases roughly linearly with the number of antennas.  Simulations show that the overhead of inverse fingerprinting is less than 2\% of that of IEEE 11ad for array sizes larger than $16\times 16$.

\section{Conclusions}

%Paragraph 1: Configuring the antennas is a source of overhead. Exploit side infomation from other bands available in many forms. Many research challenges involving SP, machine learning and communication. 

Configuring the antenna arrays, which can be done through beam training or channel estimation, is the main challenge for high data rate mmWave communications, since it introduces a significant overhead. Out-of-band information can provide a means to obtain side information about that channel without  compromising communication resources.
Some recent work  has started to develop the basic mathematical tools to establish  communication using out-of-band side information, like correlation translation between sub-6 GHz and mmWave.
Practical beam search algorithms that exploit side information, tailored to cases of particular relevance like the vehicular setting, are also being developed.
In this paper, we have described the key ideas under this original approach and we have analyzed the initial approaches to beam search that leverage side information obtained from sensors or sub-6 GHz communication systems. We have shown by simulation that the overhead reduction provided by these initial  schemes is substantial.   
  Most of the research challenges that have to be addressed to unlock the potential of out-of-band information at mmWave remain unsolved, involving  signal processing algorithms, machine learning to combine sensors and communication information, and exhaustive multi-band measurements campaigns.

\bibliographystyle{IEEEtran}
\bibliography{OOB_magazine_refs}

\end{document}